# A new method in the production of Ac-227, Ra-228, Th-228 and U-232 on Thorium target from different perspective for use in Radioisotope Power Systems and Nuclear Batteries

*Ozan Artun[1]*

Abstract | We recommended a new method from different perspective for the production of Ac-227, Ra-228, Th-228 and U-232 radioisotopes which have an important potential for use in radioisotope power systems (instead of Pu-238) and nuclear batteries, based on long half-life, high power density and low radiation level. In the method, the production of Ac-227, Ra-228, Th-228 and U-232 radioisotopes were performed by particle accelerators on Th-232 targets, instead of uranium, thorium, actinium series etc. in nuclear reactors. For this aim, to produce Ac-227, Ra-228, Th-228 and U-232 radioisotopes, we calculated cross-section and integral yields curves and simulated activities and yields of product for each reaction process under special conditions, such as during irradiation of 24 h in a particle beam current of 1 mA for the energy range of $E_{projectile}$=1000→1 MeV. The obtained data were compared with experimental EXFOR database and theoretical TENDL database in detail. The cross-section, activity and yields results of each nuclear reaction process were analyzed and discussed in terms of the suitable production of Ac-227, Ra-228, Th-228 and U-232 radioisotopes.

## 1. Introduction

Nuclear reactions provide production of new nuclei, which are not natural, to use in nuclear medicine, space researches, shielding, energy production *etc*. The nuclear reactions can be performed by nuclear reactors or particle accelerators to produce different radioisotopes in the literature [1-12]. One of the most important radioisotopes produced by nuclear reactions is energetic radioisotopes that can be used in nuclear batteries, which have vital role for microelectronic and space researches as an energy source. One of nuclear batteries used in space researches is Radioisotope Power System (RPS), which is an important technology for a spacecraft/space-probe. RPS provides heat power and electrical power for spacecraft/space-probe, particularly in deep space missions and in planetary researches [6-9,13]. It is obvious that there are mainly two practical options for ensuring a long-term source of electrical power: By exploiting the light of sun and heat obtained from a nuclear source (*e.g.* radioisotopes). These two options have some advantages and challenges, for instance, solar power is a good way to produce electricity for most Earth-orbiting spacecraft/space-probe and for certain moon missions. However, solar power is inadequate in deep space and interstellar missions, especially in harshest, darkest, coldest locations in the solar system. For example, the available sunlight at Saturn and Pluto is about 1% and 6% of what we receive at Earth. Such missions can be unfeasible or immensely restricted without the technology to reliably power space missions, such as RPS that is a type of nuclear energy technology. Heat in RPSs is used to produce electric power for operating spacecraft/space-probe systems, and RPS ensures reliable power in harsh environments as compared with solar power systems, especially in independent of changes in sunlight, temperature, charged particle radiation, or surface circumstances of spacecraft/space-probe. RPSs are part of the U.S. space program and they contribute exploration of space for more than 50 years [14]. With decades of proven success of RPSs, National Aeronautics and Space Administration (NASA) uses Radioisotope Thermoelectric Generator- abbreviated RTG- for the use of spacecraft systems and science instruments. RTGs are also called as nuclear batteries and they are not fission reactors and not the type of plutonium in nuclear weapons. At the present time, NASA uses Multi-Mission Radioisotope Thermoelectric Generator (MMRTG)[15] which is the current model of reliable RTG technology to power spacecraft (FIG. 1a), such as Perseverance and Curiosity that is on the surface of Mars. MMRTG provides electrical power by converting the heat generated by the natural radioactive decay of plutonium-238 into electricity via thermocouples in which the electricity is produced by the large difference in temperature between this hot fuel in General Purpose Heat Source (GPHS) (FIG. 1b) and the cold environment of space [15,16].

[1]*Zonguldak Bülent Ecevit University, Zonguldak, Turkey*
*e-mail: ozanartun@beun.edu.tr ; ozanartun@yahoo.com*

In addition to advantages of nuclear energy technology as compared to Solar power technology, in RTG technology, there are some challenges such as the cost of nuclear source and its production. Pu-238 nucleus has important properties such as long half-life ($T_{1/2}$=87.7 y), high power density (0.55 watt/gram) and low radiation level (majorly an alpha emitter). However, the production of Pu-238 in nuclear reactors by using uranium is quite expensive [13,14]. Instead of the use of Pu-238 in the RTG, Ac-227 ($T_{1/2}$=21.77 y, Power=2.127 watt/gram, 98.6% beta and 1.38% alpha), Ra-228 ($T_{1/2}$=5.75 y, Power= 0.016 watt/gram, >99% beta), Th-228 ($T_{1/2}$=1.91 y, Power=26.054 watt/gram; >99% alpha), and U-232 ($T_{1/2}$=68.90 y; Power=0.700 watt/gram; >99% alpha) nuclei [17] can be recommended for space-missions with RTG and for other nuclear battery types, such as alpha- and beta-voltaic batteries in microelectronic (e.g. Ra-228), based on the type of space missions, and the half-life and power density of the nuclei. Therefore, the productions of Ac-227, Ra-228, Th-228 and U-232 nuclei for use in nuclear battery technology are fairly important, and we investigated the production method from different perspective of these radioisotopes (FIG. 1c). The method for the production of Ac-227, Ra-228, Th-228 and U-232 nuclei using irradiating targets of Th-232 with proton, deuteron, triton, helium-3 and alpha particles are recommended for cheaper production via particle accelerator instead of nuclear reactors because this method directly produce the Ac-227, Ra-228, Th-228 and U-232 nuclei. This method can be chosen for instead of the production method obtained from uranium series, thorium series actinium series etc. in nuclear reactor of these radioisotopes. The thorium material Th-232 used as target in the reactions is abundant in nature as compared to the uranium material. Moreover, these four nuclei can be produced using different induced reactions via particle accelerator, e.g., Ac-227 can be produced by proton, deuteron, triton, he-3 and alpha induced reactions. Briefly, in this work, Ac-227, Ra-228, Th-228 and U-232 nuclei were produced by irradiating Th-232 with proton, deuteron, triton, helium-3 and alpha particles in the energy range between 1 MeV and 1000 MeV.

**2. Materials and Method**

To produce Ac-227, Ra-228, Th-228 and U-232 radioisotopes on Th-232 target via proton, deuteron, triton, helium-3 and alpha induced reactions in particle accelerators, we utilized from certain circumstances in the simulations of activity and yield of production in reaction processes. The target material in reaction processes is Th-232, which is stable, and the Th-232 target area is 1 cm$^2$. The purity of the target in reaction processes is >99% to obtain the best simulation results of activity and yield of production of each reaction process. On the other hands, the target thicknesses may alter based on the type of reaction process to maximize efficiency in the radioisotope productions for different reaction routes. The effective target thicknesses are given in Table 1. Also, the target materials in the simulations are of uniform thicknesses and densities during the reaction processes. The irradiation time of Th-232 for a constant beam current of 1 mA is 24 hours for all the reaction processes, and the cooling time is also 24 hours. The cross-section values for each reaction are calculated in the energy range of $E_{projectile}$=1000→1 MeV. Additionally, the cross-section calculations, activities and yields of product for the whole reaction processes are carried out by TALYS code [18]. Besides, the integral yields values are calculated by cross-section results and the mass stopping power obtained from X-PMSP program [1,19]. The calculations and simulations were performed by the conditions aforementioned above, and the obtained results were compared with experimental data obtained from EXFOR database [20] and the recommended theoretical results of obtained from the TALYS Evaluated Nuclear Data Library (TENDL) database [21].

The theoretical framework used in calculations and simulations is given by the cross-section, activity and yield equations and statements as follows:

The nuclear states in the nuclear reaction processes may be characterized by the total energy $E^{tot}$ and the total number of particles ($p$) above Fermi surface and holes ($h$) below Fermi surface where the $p$ and $h$ referred to excitons ($n$). The proton and neutron exciton numbers in the process are given by $n_x = p_x + h_x$ and $n_y = p_y + h_y$. Additionally, the charge independent particle number, the hole number refer to $p = p_x + p_y$ and $h = h_x + h_y$, respectively. Reaction stream in the exciton model during reaction process is presented by particle-hole changing (FIG. 1d). In the cross-section calculations, we utilized the two-component exciton model as follows [18]:

$$\frac{d\sigma_l^{PE}}{dE_l} = \sigma^{CF} \sum_{p_x=p_x^0}^{p_x^{max}} \sum_{p_y=p_y^0}^{p_y^{max}} W_l(p_x, h_x, p_y, h_y, E_l)\tau(p_x, h_x, p_y, h_y) \times P(p_x, h_x, p_y, h_y) \quad (1)$$

Where $\tau(p_x, h_x, p_y, h_y)$ represents lifetime for the exciton state $(p_x, h_x, p_y, h_y)$ and the lifetime $\tau$ of exciton state $(p_x, h_x, p_y, h_y)$ in Eq. (1) can be given by internal transition rates ($\lambda_x^+$ and $\lambda_y^+$), transition rate for particle-hole annihilation ($\lambda_x^-$ and $\lambda_y^-$) and the rate for transformation ($\lambda_{xy}^0(\lambda_{yx}^0)$).

$$\tau(p_x, h_x, p_y, h_y) = [\lambda_x^+(p_x, h_x, p_y, h_y) + \lambda_y^+(p_x, h_x, p_y, h_y) + \lambda_x^-(p_x, h_x, p_y, h_y) +$$
$$\lambda_y^-(p_x, h_x, p_y, h_y) + \lambda_{xy}^0(p_x, h_x, p_y, h_y) + \lambda_{yx}^0(p_x, h_x, p_y, h_y) +$$
$$W(p_x, h_x, p_y, h_y)]^{-1} \quad (2)$$

Where $W$ is the total emission rate as an integral over the whole outgoing energies and particles:

$$W(p_x, h_x, p_y, h_y) = \sum_{l=\gamma,n,p,d,t,h,\alpha} \int dE_l W_l(p_x, h_x, p_y, h_y, E_l) \quad (3)$$

$W_l$ in Eqs. (3) and (1) refers to emission rate for the emission of a particle emission energy $E_l$:

$$W_l(p_x, h_x, p_y, h_y, E_l) = \frac{2s_l + 1}{\pi^2 \hbar^3} \mu_l E_l \sigma_{l,inv}(E_l) \times \frac{\omega(p_x - Z_l, h_x, p_y - N_l, h_y, E^{tot} - E_l)}{\omega(p_x, h_x, p_y, h_y, E^{tot})} \quad (4)$$

Where $\omega$ and $E^{tot}$ are the particle-hole density and the total energy of the composite system. $\mu_l$, $Z_l(N_l)$, and $s_l$ are relative mass, the charge (neutron) number and spin for an ejectile $l$. The pre-equilibrium population $P(p_x, h_x, p_y, h_y)$ state in Eq. (1) is given by:

$$
\begin{aligned}
P(p_x, h_x, p_y, h_y) &= P(p_x - 1, h_x - 1, p_y, h_y)\Gamma_x^+(p_x - 1, h_x - 1, p_y, h_y) \\
&+ P(p_x, h_x, p_y - 1, h_y - 1)\Gamma_y^+(p_x, h_x, p_y - 1, h_y - 1) \\
&+ [P(p_x - 2, h_x - 2, p_y + 1, h_y + 1)\Gamma_x'^+(p_x - 2, h_x - 2, p_y + 1, h_y + 1) \\
&+ P(p_x - 1, h_x - 1, p_y, h_y)\Gamma_y'^+(p_x - 1, h_x - 1, p_y, h_y)] \\
&\times \Gamma_{yx}^0(p_x - 1, h_x - 1, p_y + 1, h_y + 1) \\
&+ [P(p_x, h_x, p_y - 1, h_y - 1)\Gamma_x'^+(p_x, h_x, p_y - 1, h_y - 1) \\
&+ P(p_x + 1, h_x + 1, p_y - 2, h_y - 2)\Gamma_y'^+(p_x + 1, h_x + 1, p_y - 2, h_y - 2)] \\
&\times \Gamma_{xy}^0(p_x + 1, h_x + 1, p_y - 1, h_y - 1) \quad (5)
\end{aligned}
$$

Where $\Gamma_x^+$ and $\Gamma_y^+$ are the probabilities of creating a new proton particle-hole and neutron particle-hole pair $\Gamma_x^+(\Gamma_y^+)$. Altering a proton (neutron) pair into neutron (proton) pair $\Gamma_x'^+(\Gamma_y'^+)$ is presented as follows:

$$\Gamma_x^+(p_x, h_x, p_y, h_y) = \lambda_x^+(p_x, h_x, p_y, h_y)\tau(p_x, h_x, p_y, h_y),$$

$$\Gamma_y^+(p_x, h_x, p_y, h_y) = \lambda_y^+(p_x, h_x, p_y, h_y)\tau(p_x, h_x, p_y, h_y),$$

$$\Gamma_x'^+(p_x, h_x, p_y, h_y) = \lambda_x^+(p_x, h_x, p_y, h_y)\tau'(p_x, h_x, p_y, h_y),$$

$$\Gamma_y'^+(p_x, h_x, p_y, h_y) = \lambda_y^+(p_x, h_x, p_y, h_y)\tau'(p_x, h_x, p_y, h_y),$$

$$\Gamma_{xy}^0(p_x, h_x, p_y, h_y) = \lambda_{xy}^0(p_x, h_x, p_y, h_y)\tau(p_x, h_x, p_y, h_y),$$

$$\Gamma_{yx}^0(p_x, h_x, p_y, h_y) = \lambda_{yx}^0(p_x, h_x, p_y, h_y)\tau(p_x, h_x, p_y, h_y),$$

$$\tau'(p_x, h_x, p_y, h_y), = [\lambda_x^+(p_x, h_x, p_y, h_y) + \lambda_y^+(p_x, h_x, p_y, h_y) + W(p_x, h_x, p_y, h_y)]^{-1} \quad (6)$$

Where it is clear that $\lambda_x^+$, $\lambda_y^+$, $\lambda_{xy}^o$, $\lambda_{yx}^o$ and $\tau$ refer to internal transition rates and the lifetime. If the initial condition is, $P$ is given by

$$P(p_x^0, h_x^0, p_y^0, h_y^0) = 1 \tag{7}$$

Moreover, in the determination of the activity and yield values of the reaction processes, if the target materials including various isotopes are irradiated, new isotopes may occur during irradiation via radioactive decay. Hence, when considering such targets, which involve a total of $K$ different isotopes and the number of each isotope $k$ ($N_k$), $K$ differential equations may be defined as follows:

$$\frac{dN_l(1)}{dt} = \sum_{k=1, k \neq l}^{K} \Lambda_{k \to l} N_k(t) - \Lambda_l N_l(t) \tag{8}$$

... ... ... ...

$$\frac{dN_i(t)}{dt} = \sum_{k=1, k \neq i}^{K} \Lambda_{k \to i} N_k(t) - \Lambda_i N_i(t) \tag{9}$$

... ... ... ...

$$\frac{dN_K(t)}{dt} = \sum_{k=1, k \neq K}^{K} \Lambda_{k \to K} N_k(t) - \Lambda_K N_K(t) \tag{10}$$

Where, it is obvious that the first term mentions to the feeding term where the partial formation rate for potential parent isotope to isotope $i$ ($\Lambda_{k \to i}$) is given by

$$\Lambda_{k \to i} = \lambda_{k \to i} + R_{k \to i} \tag{11}$$

Where $\lambda_{k \to i}$ is partial radioactive decay rate and $R_{k \to i}$ is partial nuclear reaction rate. On the other hand, the second term in Eq. (8) is loss term. The second term are negative values due to radioactive decay and nuclear reaction processes from isotope $i$ to any isotope. The total depletion rate for isotope $i$ ($\Lambda_i$) in Eqs. (8)-(10) can be given by total decay rate for isotope ($\lambda_i$) and total nuclear reaction rate for isotope $i$ ($R_i$):

$$\Lambda_i = \lambda_i + R_i \tag{12}$$

$$\lambda_i = \sum_{k=1, k \neq i}^{K} \lambda_{i \to k} \tag{13}$$

$$R_i = \sum_{k=1, k \neq i}^{K} R_{i \to k} \tag{14}$$

For the irradiation in target *T,* as the irradiation launches:

$$t = 0: N_T = N_T(0) \tag{15}$$

$$N_i = 0$$

$$\ldots \tag{16}$$

$$N_K = 0 \tag{17}$$

Then the loss terms can be degraded to $\Lambda_T = R_T$ for $N_T$ in Eq. (8) since the isotopes in the target materials is shaded into other isotopes during irradiation. This situation leads to beam particles that can affect other isotopes formed in target instead of original isotopes in the target. Whereas, the simulations and calculations for the reaction processes are considered for some assumptions some of which are the burning out in the target material where is relatively small, and the compositions in the target materials do not alter during the irradiation. Therefore, nuclear reaction processes or radioactive decay do not have the feeding terms for the target isotope *T*. Then, the target isotope can be given by:

$$\frac{dN_T(t)}{dt} = -R_T N_T(t) \tag{18}$$

Besides, some assumptions are valid for the isotopes formed in the irradiation *e.*g. the nuclear reaction processes do not contain loss of isotope production and form the isotopes interested. The total depletion rate for isotope *i* ($\Lambda_i$) is equal to decay rate $\lambda_i$. Thus, a new formula for the production of isotope *i* may be given by

$$\frac{dN_i(t)}{dt} = \lambda_{p \to i} N_p(t) + R_{T \to i} N_T(t) - \lambda_i N_i(t) \tag{19}$$

Where, the parent isotope *p*, which decays to isotope *i*, is given by the following equation:

$$\frac{dN_p(t)}{dt} = R_{T \to p} N_T(t) - \lambda_p N_p(t) \tag{20}$$

The solutions of $N_T$, $N_p$ and $N_i$ can be solved by the boundary conditions ($N_T(t=0) = N_T(0)$ and $N_p(t=0) = N_i(t=0) = 0$)

$$N_T(t) = N_T(0) e^{-R_T t} \tag{21}$$

$$N_p(t) = N_T(0) \frac{R_{T \to p}}{\lambda_p - R_T} \left[ e^{-R_T t} - e^{-\lambda_p t} \right] \tag{22}$$

$$N_i(t) = N_T(0) \frac{R_{T \to i}}{\lambda_i - R_T} \left[ e^{-R_T t} - e^{-\lambda_i t} \right] + N_T(0) \frac{R_{T \to p} \lambda_{p \to i}}{\lambda_p - R_T} \left[ \frac{e^{-R_T t} - e^{-\lambda_i t}}{\lambda_i - R_T} - \frac{e^{-\lambda_p t} - e^{-\lambda_i t}}{\lambda_i - \lambda_p} \right] \tag{23}$$

Additionally, the maximal yield in the reaction processes is obtained by the irradiation time which is produced by setting the derivative of $N_i(t)$ to zero:

$$t_{max} = \frac{\ln(\lambda_i / R_T)}{\lambda_i - R_T} \tag{24}$$

Where $t_{max}$ depends on $\lambda_i (ln2/T_i^{1/2})$ of isotope $i$ and $R_T$. If the initial conditions of $N_i(t)$ are applied in Eq. (9), number of targets atoms at $t=0$ ($N_i(0)$) may be found by the following equation:

$$N_i(0) = \frac{N_A}{M} B_i \rho V_{tar} \tag{25}$$

Where, $M$ is the mass number and $N_A$ is Avogadro's number. $\rho$ and $B_i$ refer to the mass density (in g/cm$^3$) and the abundance of isotope $i$, respectively. In Eq. (25), the active target volume $V_{tar}$ (in cm$^3$) is given by

$$V_{tar} = S_{beam} \int_{E_{back}}^{E_{beam}} \left( \frac{dE}{dx} \right)^{-1} dE \tag{26}$$

Where $S_{beam}$ is the product of the beam surface in cm$^2$ and $(dE/dx)$ is the stopping power of the target material. The integration limits $E_{beam}$ and $E_{back}$ are fixed by ensured projectile energy range within the target fixed by cross-section as a function of projectile energy.

The production rate $R_{T \to i}$, which a beam hits the target, will become in atoms/s of isotope $i$ for the nuclear reaction on the target $T$ and it can be defined by the electron charge $q_e$, beam current $I_{beam}$ (in A) and the residual production cross-section of $i$ ($\sigma_i^{rp}(E)$) in mb:

$$R_{T \to i} = \frac{I_{beam}}{z_p q_e} \frac{1}{V_{tar}} \int_{E_{back}}^{E_{beam}} \left(\frac{dE}{dx}\right)^{-1} \sigma_i^{rp}(E) dE \qquad (27)$$

Where the statement $I_{beam}/z_p q_e$ refers the number of projectiles knocking on the target per second. On the other hand, for target $T$, the production rate $R_T$ in s$^{-1}$ including the whole reaction channels is defined by difference between the non-elastic cross-section $\sigma_{non}(E)$ and inelastic cross-section $\sigma_{in}(E)$ of isotope $i$ (in mb).

$$R_T = \frac{I_{beam}}{z_p q_e} \frac{1}{V_{tar}} \int_{E_{back}}^{E_{beam}} \left(\frac{dE}{dx}\right)^{-1} (\sigma_{non}(E) - \sigma_{in}(E)) dE \qquad (28)$$

The activity statement is given for the produced isotope $i$ is given by:

$$A_i(t) = \lambda_i N_i(t) \qquad (29)$$

$$N_i(t) = N_T(0) R_{T \to i} t \qquad (30)$$

Then, the activity can be given as follows:

$$A_i(t) = \lambda_i N_T(0) R_{T \to i} t, \qquad (31)$$

Additionally, based on irradiation time $t$ and the production rate $R_{T \to i}$, the product yield stated in MBq/mAh is described by an exclusive irradiation time $t$ and beam current $I$ [18]:

$$Y_i(t) = \frac{A_i(t)}{R_{T \to i} t} \qquad (32)$$

On the other hands, the integral yield results of the each reaction process can be given by the activation equation as dependent on atomic weight $M$, the projectile current $I_{beam}$, Avogadro constant $N_A$ and isotopic abundance $H$, the cross-section of the reactions $\sigma(E)$, the decay constant ($\lambda = ln2/T_{1/2}$), the proton, deuteron, triton, helium-3 and alpha mass stopping powers of Th-232 target obtained by the X-PMSP 2.0 program [1,4,5,7-9,19].

$$Y = \frac{N_A H}{M} I_{beam}(1 - e^{-\lambda t}) \int_{E_1}^{E_2} \frac{\sigma(E)dE}{\left(\frac{dE}{d(\rho x)}\right)} \tag{33}$$

## 3. Results and discussions

### 3.1. Cross-section calculations of reaction processes

To produce Ac-227, Ra-228, Th-228 and U-232 on Th-232 targets via particle accelerators, the cross-section calculations were calculated for the proton, deuteron, triton, helium-3 and alpha-induced reaction processes, and the calculated cross-section curves were compared with the experimental data in the literature as shown in FIG. 2. Each figure dependents on incident particle energy based on induced reaction type. Taken into account the experimental data in the literature, there is the lack of cross-section data where are only proton- and alpha-induced reactions. For the deuteron, triton and helium-3-induced reactions, no experimental data is not available for the production of Ac-227, Ra-228, Th-228 and U-232. For this aim, we added the cross-section data in TENDL-2019 nuclear data library, where the projectile energies are maximally up to 200 MeV, to the cross-section figures. The cross-section calculations for each reaction were performed in the energy region between 1 MeV and 1000 MeV. It is obviously presented in FIG. 2, where the production of Ac-227, Ra-228, Th-228 and U-232 radioisotopes includes few experimental data reported by Griswold et al (2016), Ermoloev et al. (2012), Gauvin (1963), Lefort et al. (1961) [22-25] for the production of Ac-227, Lefort et al. (1961) for the production of Ra-228, and Hogan et al. (1979) [26], Lefort et al. (1961) in the proton-induced reaction. In addition to proton-induced reaction on Th-232 target, the data reported by ForemanJr et al. (1959) [27] in the production of $^{232}$U via alpha-induced reaction process were presented in FIG. 2e as a function of the incident energy. Ac-227, Ra-228 and Th-228 radioisotopes may be produced by Th-232(p,x) reaction process where the radioisotope Th-228 has the highest cross-section value ~142 mb at 75 MeV, and the cross-section curves of Ac-227 and Ra-228 radioisotopes reach ~26 mb at 100 MeV and ~0.7 mb at 165 MeV, respectively. In the Th-232(p,x) reaction process, the calculated cross-section curve for the production of Ac-227 overlaps with the experimental data up to 55 MeV. Beyond 55 MeV, the experimental data deviate from the calculated cross-section, and Griswold et al.'s (2016) one data point intersects the cross-section curve at ~130 MeV. Moreover, the calculated results are also consistent with the TENDL results. The calculated results are even closer to the experimental results than the TENDL results, especially for the production of Th-228 and Ac-227 radioisotopes. For the production of Th-228, the experimental data and the calculated results demonstrated that the maximum cross-section values are in the energy region between 45 MeV and 90 MeV where the experimental data have cross-section values than that of the calculated result. In the Ra-228, there is only a data point at 150 MeV and the experimental data is not adequate to reliably compare with the calculated cross-section curve. On the other hand, in the Th-232($\alpha$,x) reaction process, two data points reported by ForemanJr et al. (1959) for the production of U-232 are consistent with the calculated cross-section curves and the TENDL results in each point of FIG. 2e. The cross-section value of U-232 achieves to 195 mb at 38 MeV. When taken into account the cross-section curves of the radioisotopes in FIG. 2, it is noted here that the reaction types are very important in the production of Ac-227, Ra-228, Th-228 and U-232 radioisotopes since reaction type provides advantage if suitable reaction process is considered. For instance, in the production of Ac-227, the triton-induced reaction has the biggest cross-section values as compared to the other reaction processes. Similarly, for the production of U-232, which can be produced by he-3- and alpha-induced reaction processes, the alpha-induced reaction clearly has advantage when considered cross-section value and incident particle energy. For the Th-228, the cross-section curve of deuteron induced reaction ascend to 272 mb at 83 MeV. Except for the radioisotope Th-228, the he-3-induced reaction (FIG. 2d) is not suitable for the production of Ac-227, Ra-228, and U-232 radioisotopes. In general, the cross-section curves of these four radioisotopes arrive to maximum values in energy region between 50 and 150 MeV, and the harmony between the TENDL and the calculated results is suitable. However, the TENDL results have higher cross-section values than those of the calculated results, except for the production of U-232. The maximum cross-section values and energy values are presented in Table 1 in detail. The cross-section curve of the reaction processes is inadequate to recommend the appropriate production pathway of the radioisotope production. For this aim, we simulated the activity and yield of product of each reaction process in 1 mA current beam during irradiation time of 24 h for the thick target Th-232.

### 3.2. Activities and Yields of Product of Reaction Processes

The activity and yield of product curves for the productions of Ac-227, Ra-228, Th-228 and U-232 radioisotopes in particle accelerators are demonstrated in FIG. 3 and FIG. 4 for the beam current of 1 mA and irradiation time of 24 h. The processes include the proton, deuteron, triton, helium-3 and alpha-induced reactions in the energy range of $E_{projectile}=1000 \rightarrow 1$ MeV. Under the circumstances mentioned in section 2, the activities and yields of product for each reaction process are simulated by TALYS code. Contrary to the cross-section curves, there are differences in activity and yield results such as the production of Th-228 which is the biggest activity and yield values among the whole reaction processes because the Th-228 radioisotope is another isotope of target material Th-232. In the Th-232($\alpha$,x) reaction process, the production of U-232 has lower activity and yield values than the other radioisotopes unlike the cross-section curve. Nevertheless, the alpha-induced reaction process to produce U-232 is more consistent than he-3-induced reaction process where the activity arrives ~1 MBq at $7^{th}$ hour. When considered the appropriate reaction processes in the production of Ac-227, Ra-228, Th-228 and U-232 radioisotopes: The production of Th-228 has maximum activity and yield values in proton induced reaction process, ~$5.3 \times 10^5$ MBq and ~$2.3 \times 10^4$ MBq/mA, respectively. It is important to note here that this is advantage for the production of Th-228 since the triton-induced reaction process is more expensive proton- and deuteron-induced reaction processes due to cost of obtaining triton particle in the world. Another important point, in the production of Ac-227 and Ra-228 radioisotopes, the proton- and deuteron-induced reactions give more convenient activity and yield results than those of the other reactions, even if the cross-section curve of Ac-227 radioisotope arrives the maximum value (~383 mb) in the triton induced reaction process (Th-232(t,x)). The maximum activity and yield results of each reaction process are given in Table 1 for the irradiation time at 24 hours. It has been noted that the obtained activity and yield values of all reaction processes are consistent with the cross-section results. The highest activity and yield values during irradiation of the other reactions are shown in Table 1. On the other hands, we considered the activity, yield of product and cross-section curves of reaction processes for the production of Ac-227, Ra-228, Th-228 and U-232 radioisotopes. While the certain reaction types can give suitable cross-section curves, the activity and yield of product results for the production of the same radioisotope in the same reaction processes do not demonstrate the same success. Hence, we calculated the integral yield results of the whole reaction processes using the cross-section curves and the mass stopping power of Th-232 target and the integral yield results were presented in FIG. 5 as a function of projectile energy. The integral yield results help to determine the suitable production energy ranges of the reaction processes.

### 3.3. Integral Yield Calculations of Reaction Processes

To estimate the integral yields of (p,x), (d,x), (t,x), (he-3,x) and ($\alpha$,x) reactions on Th-232 target, we used the cross-section results obtained from FIG. 2, and the mass stopping powers of Th-232 target material for p, d, t, he-3 and $\alpha$ particles are calculated by the X-PMSP program. The integral yield calculations for the production of Ac-227, Ra-228, Th-228 and U-232 nuclei via

Th-232(p,x), Th-232(d,x), Th-232(t,x), Th-232(he-3,x) and Th-232(α,x) reaction processes have been performed in FIG. 5 for the irradiation time of 24 h, the constant beam current of 1 mA and the energy range of 1-1000 MeV. In general, the integral yields value of the reaction processes increase as the projectile energy increase, except for Th-232(α,x)U-232 and Th-232(t,x)Ac-227 reactions, where the formation of the radioisotopes has occurred up to 200 MeV. Figure 5 clearly shows that Ac-227 forms about 90% up to 200 MeV of triton incident energy. The other reactions in the production of Ac-227 involve extensive energy range. The integral yield curves of the proton- and deuteron-induced reactions reach ~11.51x10$^3$ GBq/mAh and ~3.07x10$^3$ GBq/mAh, respectively. The integral yield value for Th-232(d,x)Th-228 reaction is 6647.38 GBq/mAh, and it is evidently higher than Th-232(t,x)Th-228 (2108.09 GBq/mAh), Th-232(he-3,x)Th-228 (285.58 GBq/mAh) and Th-232(α,x)Th-228 (124.48 GBq/mAh) reactions. For the production of Th-228, we can say that Th-232(d,x)Th-228 reaction has more appropriate conditions than those of Th-232(t,x)Th-228, Th-232(he-3,x)Th-228, and Th-232(α,x)Th-228 reactions. The appropriate energy range for the production of Th-228 in Th-232(d,x)Th-228 reaction is from 55 MeV to 180 MeV deuteron incident energy. Moreover, it has been noted that the formation of U-232 in Th-232(α,x) reaction makes up about 95% up to 100 MeV alpha incident energy. In this reaction, the suitable energy range for the production of U-232 is between 40 MeV and 85 MeV. Similar situations are also available for the other reactions, and the appropriate energy ranges for the production of Ac-227, Ra-228, Th-228 and U-232 in each reaction process are presented in Table 1.

## 4. Conclusion

The estimation of Ac-227, Ra-228, Th-228 and U-232 radioisotopes that might be utilized as an energy source in RTGs and nuclear batteries in particle accelerators were investigated in the energy range E$_{projectile}$=1000→1 MeV. For the production of Ac-227, Ra-228, Th-228 and U-232 radioisotopes, the Th-232 targets are bombarded with proton, deuteron, triton, he-3 and alpha particles during irradiation time of 24 h with constant beam current of 1 mA. The calculated cross-section curves, the simulated activity and yield of product curves for each reaction process were determined. The obtained data were compared with the experimental data obtained from EXFOR database and theoretical TENDL library database. It was obvious that the obtained data agreed with the experimental data and TENDL results. However, there was no experimental data in the literature for each reaction process. Additionally, TENDL results involve only the cross-section curves of reactions in the energy region between 1 MeV and 200 MeV where the calculated data were more consistent than TENDL results.

In the production of Ac-227, Ra-228, Th-228 and U-232 radioisotopes, the proton- and deuteron-induced reaction processes on Th-232 targets demonstrated suitable results for the productions of Ac-227 and Th-228 radioisotopes when considered the cross-section, activity and yield curves, especially in Th-232(d,x)Th-228 reaction, where the cross-section curve reached ~272 mb. In the triton induced reaction process, the maximum cross-section values for the production of Ac-227 was ~383 mb. Notably, Th-232($\alpha$,x) reaction process can be recommended for the production of U-232, which the suitable energy range is 85→35, and the U-232 has the maximum cross-section value at 38 MeV. Ra-228 was produced in trace amount and the production of Ra-228 could be proposed the proton- and deuteron-induced reaction processes. It is obvious that the production of Ac-227, Ra-228, Th-228 and U-232 radioisotopes can be carried out directly by the charged-induced reaction processes on Th-232 targets in particle accelerators. Moreover, it is important to note here that Ac-227, Ra-228, Th-228 and U-232 radioisotopes as alpha and beta sources can be proposed for use in different nuclear batteries such as RTGs, and beta/alpha-voltaic nuclear batteries for low power electronic devices since these radioisotopes have suitable properties.

**FIGURES:**

**Figure 1. (a)** A cross-sectional view of MMRTG. (**b**) A cross-sectional view of GPHS. (**c**) The productions of Ac-227, Ra-228, Th-228 and U-232 radioisotopes on Th-232 target via proton, deuteron, triton, helium-3 and alpha-induced reaction processes. (**d**) The particle-hole changing in the exciton model during reaction process.

**Figure 2.** The calculated, recommended and experimental cross-section results for Ac-227, Ra-228, Th-228 and U-232 radioisotopes.

**Figure 3.** The simulated activity curves of Ac-227, Ra-228, Th-228 and U-232 radioisotopes for the reaction processes.

**Figure 4.** The simulated yield of product curves of Ac-227, Ra-228, Th-228 and U-232 radioisotopes for the reaction processes.

**Figure 5.** The calculated integral yield curves of Ac-227, Ra-228, Th-228 and U-232 radioisotopes for the reaction processes.

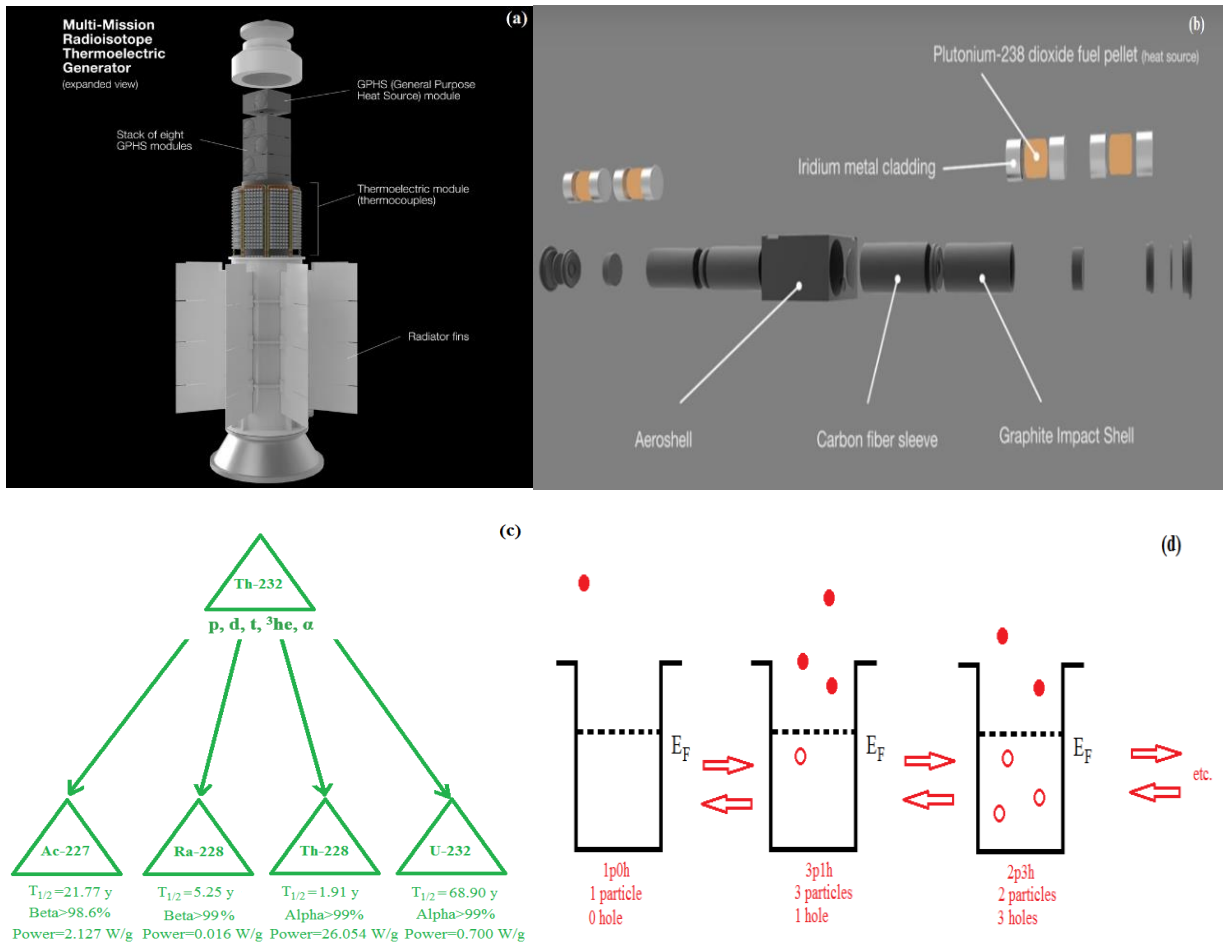

**Figure 1.**

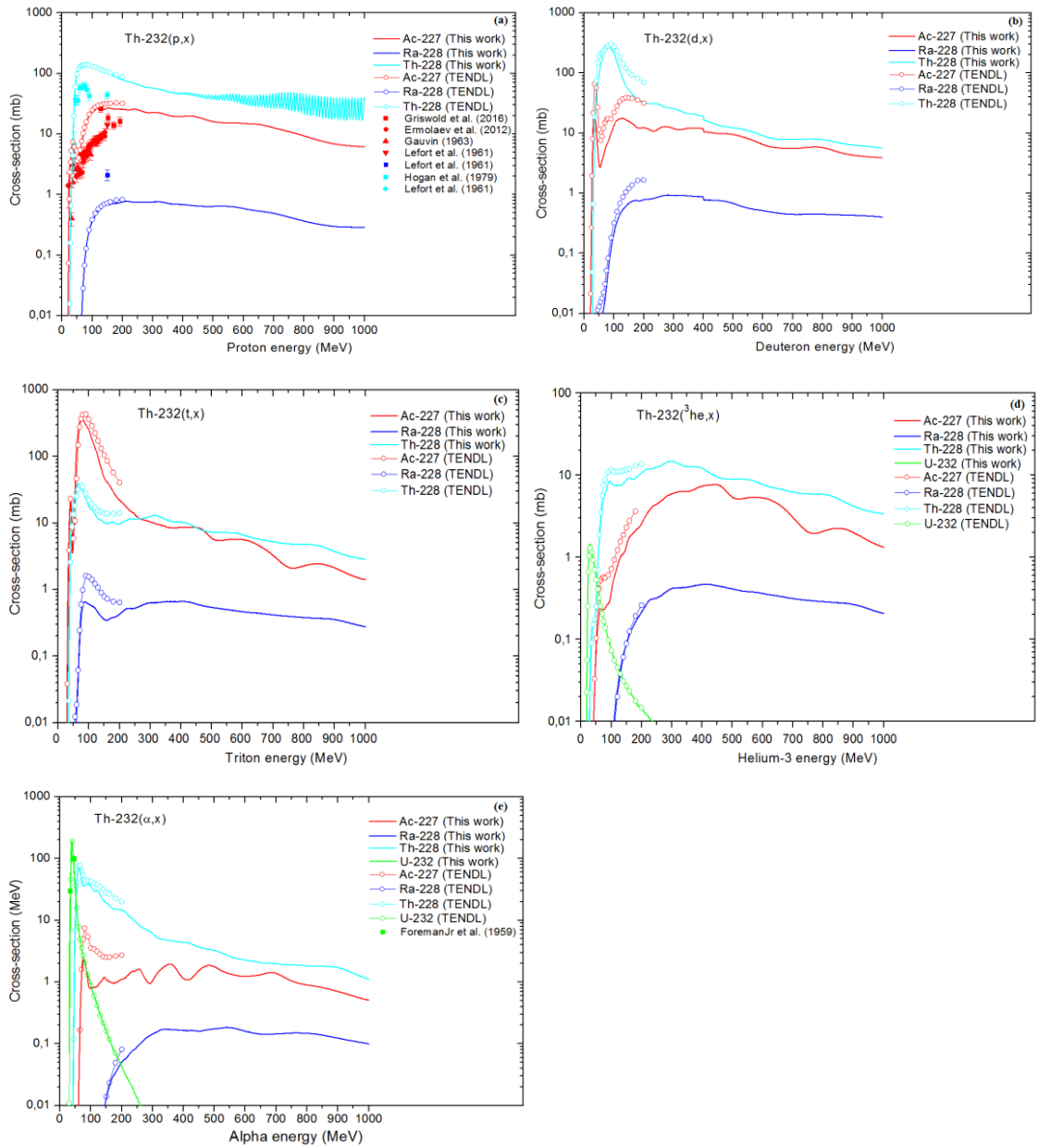

**Figure 2.**

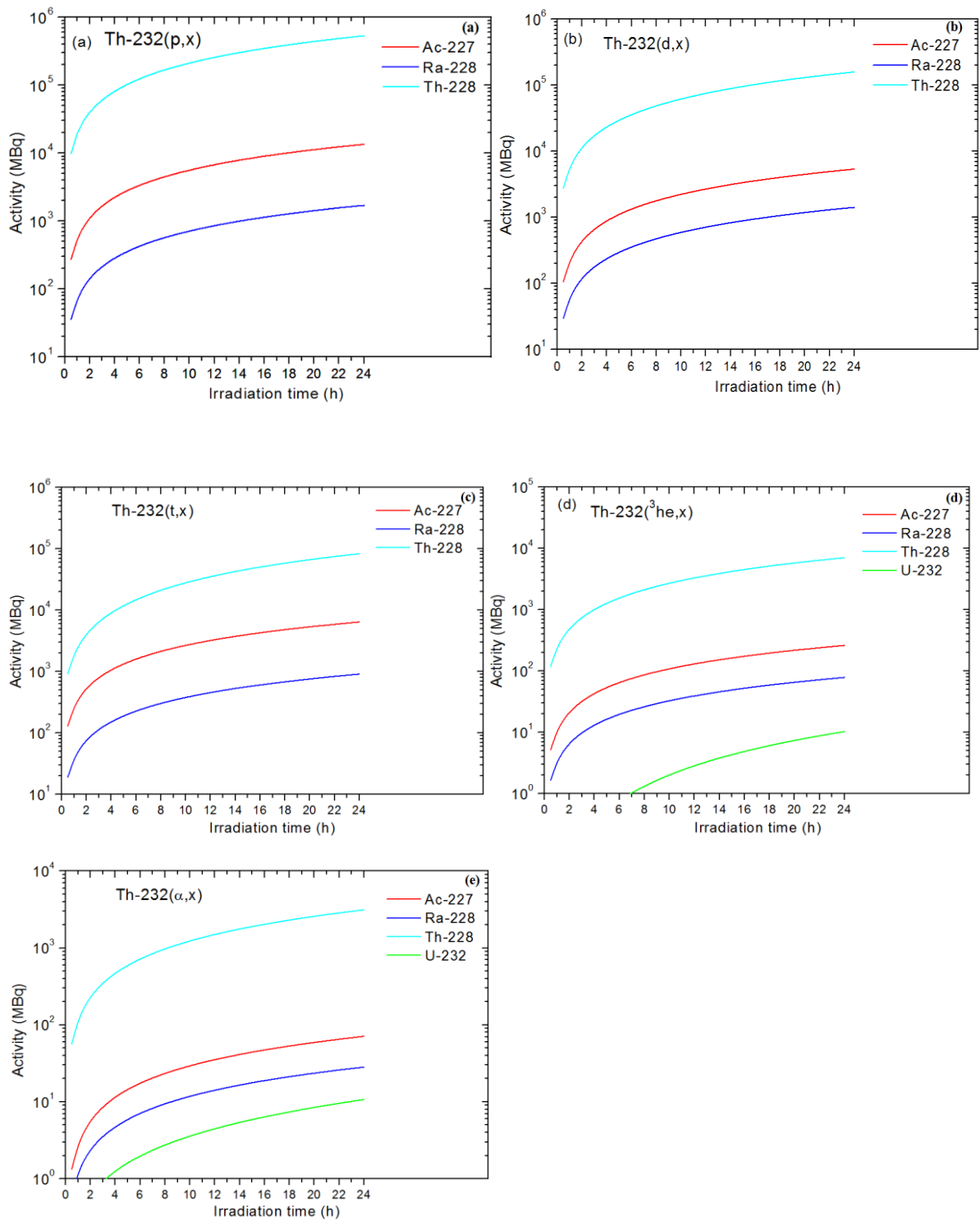

**Figure 3.**

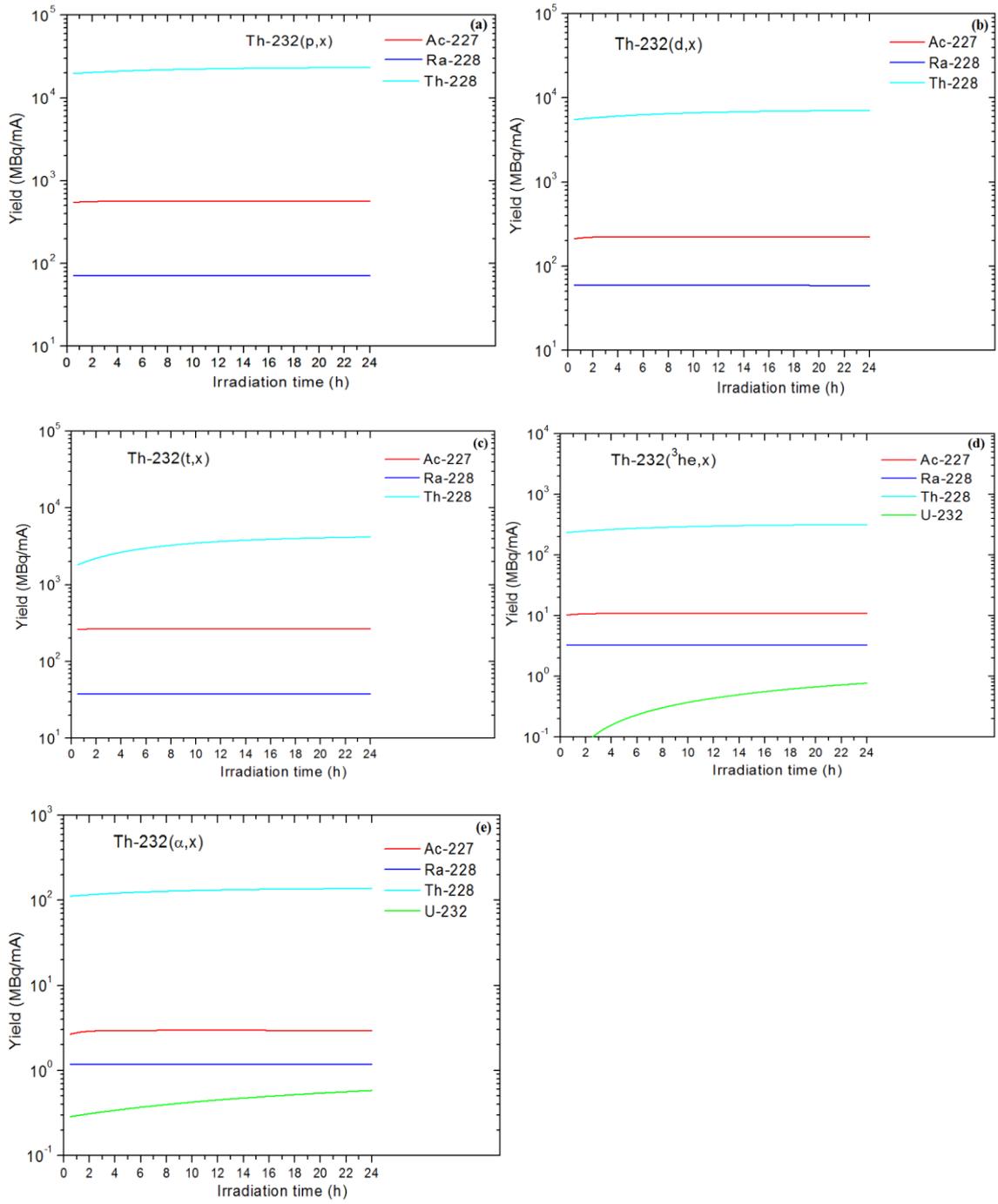

**Figure 4.**

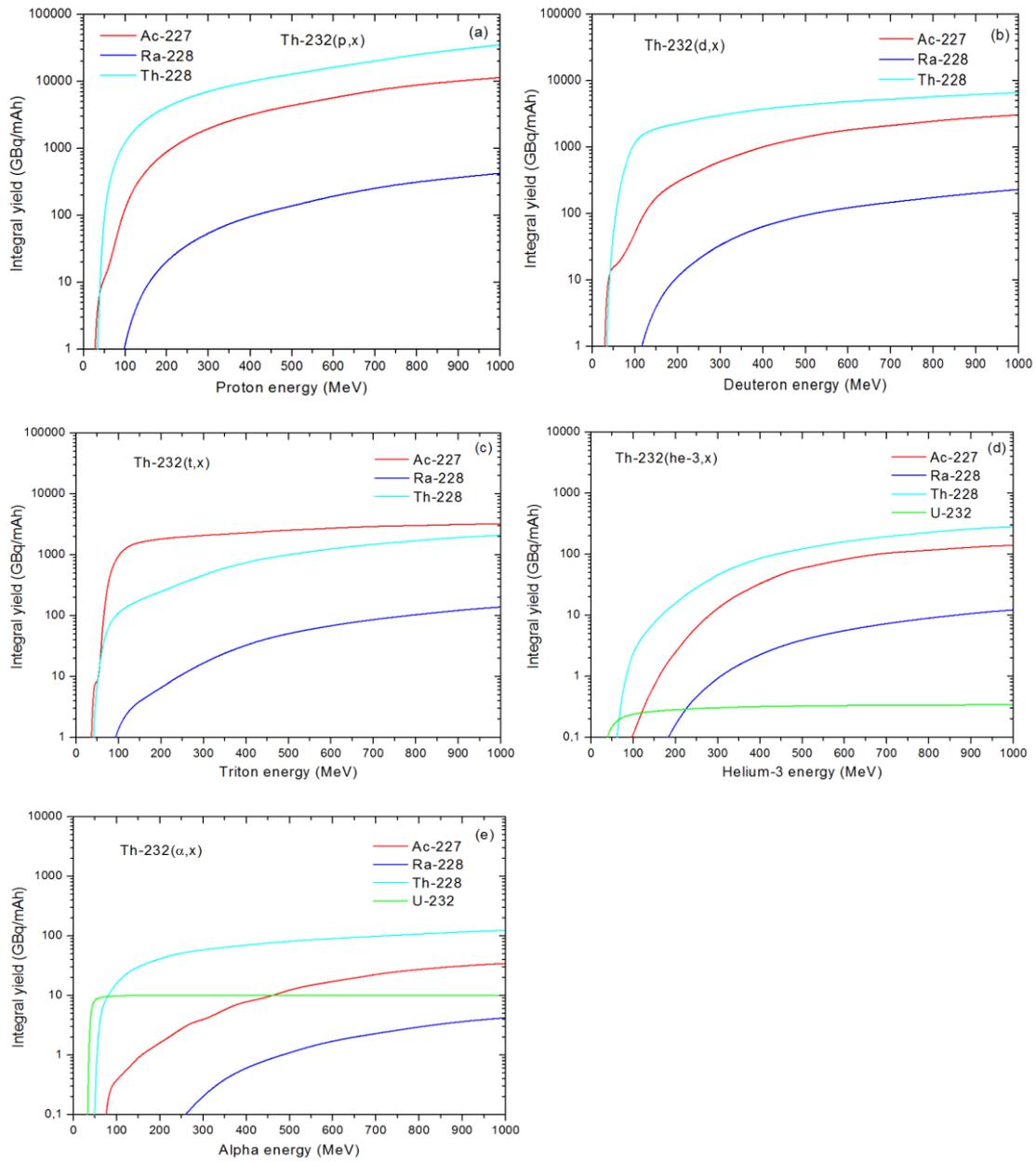

**Figure 5.**

**TABLE:**

**Table 1.** The obtained data for the production of Ac-227, Ra-228, Th-228 and U-232 radioisotopes.

| Target | Reactions | Effective thick target thickness (cm) | Products | Maximum cross-section (mb) | Energy value at maximum cross-section (MeV) | Activity at 24 h (MBq) | Suitable energy range (MeV) | Integral yield at 1000 MeV (GBq/mAh) |
|---|---|---|---|---|---|---|---|---|
| Th-232 | (p,x) | 54.374 | Ac-227 | 26.84 | 157.00 | 13542.30 | 350→100 | 11516.94 |
| | | | Ra-228 | 0.78 | 213.00 | 1712.74 | 330→170 | 429.43 |
| | | | Th-228 | 142.00 | 76.00 | 534788.00 | 270→60 | 36328.82 |
| | (d,x) | 40.122 | Ac-227 | 22.38 | 35.00 | 5381.29 | 350→115 | 3076.58 |
| | | | Ra-228 | 0.92 | 289.00 | 1418.71 | 500→195 | 232.54 |
| | | | Th-228 | 272.36 | 83.00 | 159520.00 | 180→55 | 6647.38 |
| | (t,x) | 32.023 | Ac-227 | 383.23 | 75.00 | 6438.02 | 175→80 | 3252.58 |
| | | | Ra-228 | 0.65 | 88.00 | 909.44 | 625→290 | 139.61 |
| | | | Th-228 | 38.70 | 64.00 | 83295.30 | 420→65 | 2108.09 |
| | (he-3,x) | 8.006 | Ac-227 | 7.66 | 438.00 | 261.86 | 510→175 | 141.50 |
| | | | Ra-228 | 0.47 | 415.00 | 78.52 | 600→245 | 12.31 |
| | | | Th-228 | 14.77 | 300.00 | 7063.92 | 510→100 | 285.58 |
| | | | U-232 | 1.47 | 31.00 | 10.29 | 120→50 | 0.34 |
| | (α,x) | 6.724 | Ac-227 | 2.31 | 78.00 | 71.01 | 710→95 | 34.74 |
| | | | Ra-228 | 0.171 | 352.00 | 28.28 | 810→354 | 4.26 |
| | | | Th-228 | 71.76 | 61.00 | 3127.62 | 300→65 | 124.48 |
| | | | U-232 | 194.96 | 38.00 | 10.72 | 85→35 | 10.22 |